%% file: main.tex
\documentclass[sigconf]{acmart}

\usepackage[english]{babel}
\usepackage[utf8]{inputenc}
\usepackage{ulem}
\usepackage{amsmath,mathtools}
\usepackage[noend]{algpseudocode}
\usepackage{subcaption}
\usepackage[ruled,vlined,linesnumbered,noresetcount,noend]{algorithm2e}
\usepackage{makecell}

\usepackage{enumitem}
\usepackage{xspace}

\newcommand{\eat}[1]{}
\newcommand{\ChooseSubtree}{\mbox{$\mathsf{ChooseSubtree}$}\xspace}
\newcommand{\Split}{\mbox{$\mathsf{Split}$}\xspace}

\AtBeginDocument{%
  \providecommand\BibTeX{{%
    \normalfont B\kern-0.5em{\scshape i\kern-0.25em b}\kern-0.8em\TeX}}}

\begin{document}

\title{A Reinforcement Learning Based R-Tree for Spatial Data Indexing in Dynamic Environments}


\author{Tu Gu$^{1}$, Kaiyu Feng$^{1}$, Gao Cong$^{1}$, Cheng Long$^{1}$, Zheng Wang$^{1}$, Sheng Wang$^{2}$}

\affiliation{
  \institution{$^1$School of Computer Science and Engineering, Nanyang Technological University, Singapore}
   \institution{$^2$DAMO Academy, Alibaba Group}
}

\email{gutu0001@e.ntu.edu.sg, {kyfeng, gaocong, c.long, wang_zheng}@ntu.edu.sg, sh.wang@alibaba-inc.com}


\begin{abstract}
  
  Learned indices have been proposed to replace classic index structures like B-Tree with machine learning (ML) models. They require to replace both the indices and query processing algorithms currently deployed by the databases, and such a radical departure is likely to encounter challenges and obstacles. In contrast, we propose a fundamentally different way of using ML techniques to improve on the query performance of the classic R-Tree without the need of changing its structure or query processing algorithms. Specifically, we develop reinforcement learning (RL) based models to decide how to choose a subtree for insertion and how to split a node when building an R-Tree, instead of relying on hand-crafted heuristic rules currently used by  R-Tree and its variants. Experiments on real and synthetic datasets with up to more than 100 million spatial objects clearly show that our RL based index outperforms R-Tree and its variants in terms of query processing time. 
  
\end{abstract}

\maketitle

\input{Introduction}
\vspace*{-2mm}
\input{problem.tex} 

\vspace*{-2mm}
\input{method}

\vspace*{-2mm}
\section{Experiments} \label{Experiments}


\subsection{Experimental Setup} \label{Experimental Setup}

\indent \textbf{Datasets.} We use 3 synthetic datasets (1-3) used in previous work on spatial indices \cite{arge2008priorityrtree, qi2018parallelizability, qi2020deeplearningrtree}, and 2 large real-life datasets (4-5).

\begin{enumerate}[leftmargin=*]
    \item Skew (SKE): It consists of small squares of a fixed size. The $x$ and $y$ coordinates of the squares centers are randomly generated from a uniform distribution in the range $[0,1]$ and then “squeezed” in the $y$-dimension, that is, each square center $(x,y)$ is replaced by $(x,y^c)$ where $c$ is the skewness and the default value of which is set to be 9;
    \item Gaussian (GAU): It consists of small squares of a fixed size. The coordinates of the center of a square are $(x,y)$ where $x$ and $y$ are randomly generated from a Gaussian distribution with mean $\mu = 0.5$ and standard deviation $\sigma = 0.2$;
    \item Uniform (UNI): It consists of small squares of a fixed size. The $x$ and $y$ coordinates of the centers of the squares are randomly generated from a uniform distribution in the range $[0,1]$;
    \item OSM China (CHI): It contains more than 98 million locations in China extracted from OpenStreetMap;
    \item OSM India (IND): It contains more than 100 million locations in India extracted from OpenStreetMap.
\end{enumerate}


Note that the centers of all spatial data objects in synthetic datasets fall within the unit square.

\noindent \textbf{Queries.} For model training, we run range queries over both the reference R-Tree and the RLR-Tree. Range queries of different sizes are generated. When a range query is generated, we first set its center the same as the center of the last inserted object. Its length to width ratio is then randomly selected in the range $[0.1, 10]$. To compare the query performance of RLR-Tree with others, 
we randomly generate 1,000 range queries 
for each query size ranging from 0.005\% to 2\% of the whole region. The testing queries sizes follow the setting in previous work \cite{qi2018parallelizability}. Note that queries used for training and testing are generated separately and hence different.

\noindent \textbf{Baselines.} 
We compare with R-Tree and its variants that are designed for dynamic environments where updates occur 
frequently. 
Baselines used in the experiments include R-Tree \cite{guttman1984rtree}, which is also the reference tree used for model training, R*-Tree \cite{beckmann1990rstar}, RR*-Tree \cite{beckmann2009revisedrstartree} which is reported to have the best query performance among R-Tree variants built using one-by-one insertion for supporting queries in dynamic environments.
We also compare with LISA \cite{li2020lisa} which is the only disk based learned index that 
returns exact results for range queries and KNN queries. Note that LISA only supports point data, so it is tested on CHI and IND datasets only.
%
We do not compare with packing R-Trees that are designed for the static databases~\cite{kamel1993packing} or other learned indices as 
they are not designed for dynamic environments as discussed in Section~\ref{Related Work}.

\noindent \textbf{Measurements.} 
%
For measurements of query performance, we consider both running time and the I/O cost. We find that both measures yield qualitatively consistent results (as exemplified in Figure~\ref{RL ChooseSubtree Query Processing Time}).
%
We  follow \cite{li2020lisa} to report the average relative I/O cost mainly. For each query, the relative I/O cost of an index is computed by the ratio of the I/O cost for it to answer the query to the I/O cost for an R-Tree to answer the same query. 
Smaller relative I/O costs indicate better query performance compared with the R-Tree. 
%

\noindent \textbf{Parameter settings.} Table \ref{Parameters and Values} shows a list of parameters and their corresponding values tested in our experiments.  The default settings are bold. For all R-Tree variants evaluated in this paper, we maintain a maximum of 50 and a minimum of 20 child nodes per tree node.

\begin{table}[tbh]
\normalsize
\vspace*{-3mm}
\centering
\caption{Parameters and Values}
\label{Parameters and Values}
\vspace{-3ex}
\begin{tabular}{| l | l |}
\hline
\textbf{Parameters} & \textbf{Values}  \\
\hline
\hline
Data distribution & SKE, \textbf{GAU}, UNI   \\
Dataset size (million) & 1, 5, 10, \textbf{20}, 100 \\
Training set size (thousand) & 25, 50, \textbf{100}, 200 \\
Training query size (\%) & 0.005, \textbf{0.01}, 2 \\
Testing query size (\%) & 0.005, \textbf{0.01}, 0.05, 0.1, 0.5, 1, 2 \\
Action space size $k$ & \textbf{2}, 3, 5, 10 \\
Number of dimensions & \textbf{2}, 4, 6, 8, 10 \\
\hline
\end{tabular}
\vspace*{-3mm}
\end{table}

The DQN models for both \ChooseSubtree and \Split contain 1 hidden layer of 64 neurons with SELU \cite{klambauer2017selfSELU} as the activation function. In the training process, the learning rate is set to be 0.003 for RL \ChooseSubtree and 0.01 for RL \Split. The initial value of $\epsilon$ is set to be 1 and the decay rate is set to be 0.99. The value of $\epsilon$ is never allowed to be less than 0.1 in order to maintain a certain degree of exploration throughout model training. The replay memory can contain at most 5,000 $(s, a, r, s')$ tuples. Network update is done by first sampling a batch of 64 tuples from the replay memory. Then $\Theta$ is updated by using gradient descent of the MSE loss function to close the gap between the Q-value predicted by $\Theta$ and the optimal Q-value derived from $\Theta^-$. The discount factor is set to be 0.95 for RL \ChooseSubtree and 0.8 for RL \Split. Synchronization of $\Theta^-$ with $\Theta$ is done once every 30 network updates.

During the model training for RL \ChooseSubtree (resp. RL \Split), the deterministic splitting (resp. insertion) rules are set to be the same as that used by the reference tree which is minimum overlap partition (resp. minimum node area enlargement). 
We train the RL \ChooseSubtree and \Split models for 20 and 15 epochs, respectively, and set $parts$ in Algorithm \ref{Splitting Model Training} to be 15, i.e.,  the training dataset is divided into 15 equal parts.
 %
The action space size $k$ for both RL \ChooseSubtree and RL Split is set to be 2 by default. Note that the trivial case of $k=1$ simply gives us the reference tree. 
 
We train our models on NVIDIA Tesla V100 SXM2 16 GB GPU using PyTorch 1.3.1. All indices 
are coded using C++. 

\vspace{-2mm}
\subsection{Experimental Results} \label{Experimental Results}


Our experiments aim to find out:

\begin{enumerate}
    \item Can RL \ChooseSubtree and RL \Split individually build better R-Trees (Section~\ref{RL Insertion Experiments} and Section~\ref{RL Splitting})?
    
    \item Can RLR-Tree outperform the baselines for range queries and KNN queries (Section~\ref{Combined RL Model})?
    
    \item How well RLR-Tree handles dynamic updates considering changes in data distributions (Section~\ref{dynamic updates})?
    
    \item The effect of different parameters on performance such as the training dataset size, the action space size of the RL models and the training query size (Section~\ref{subsed:exp:parameter});
    
    \item How well RLR-Tree scales with dimensions and how large RLR-Tree's construction time and size are (Section~\ref{subsec:other exp})?
\end{enumerate}

\vspace{-2mm}
\subsubsection{RL \ChooseSubtree.} \label{RL Insertion Experiments}
\begin{figure}[tbh]
\vspace*{-3mm}
\begin{center}
\noindent
  \includegraphics[width=0.85\linewidth]{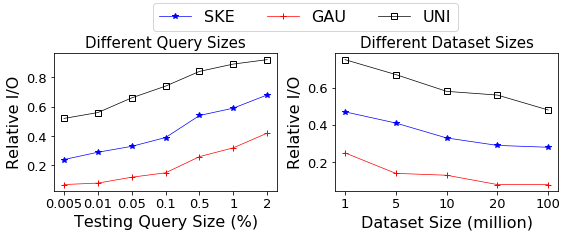}
  \end{center}
  \vspace*{-5mm}
    \caption{Performance of RL \ChooseSubtree}
    \label{RL insertion model performance}
\vspace*{-2mm}
\end{figure}

Figure \ref{RL insertion model performance} reports the average relative I/O cost of the RL \ChooseSubtree on the 3 synthetic datasets by varying the query region size and dataset size, respectively.
RL \ChooseSubtree  outperforms R-Tree consistently over different query sizes on all datasets. The best relative I/O cost is 0.07 and observed on GAU dataset for query size 0.005\%.  
%
%
We also observe that the performance of RL \ChooseSubtree gets better as query size decreases. This is because a query with a larger size intersects with a larger number of tree nodes, and more tree nodes will then be traversed when answering such a query. 
In this case, all R-Tree based indices need to visit a 
larger portion of the data and the difference between different indexing techniques will diminish. 
Consider an extreme case where a range query covers the entire data space. Then all tree nodes will be traversed when answering the query, irrespective the index used, and thus relative I/O cost will be close to 1.


The RL \ChooseSubtree model also outperforms the R-Tree consistently over different dataset sizes.
It is remarkable that although the training dataset size is only 100,000, the trained model can be applied on large datasets successfully.
Furthermore, the model performance gets better on larger datasets! 
This could be because 
as dataset gets larger the index tree becomes larger and more RL based subtree selections are done on average. Hence, the accumulated benefits from the RL based \ChooseSubtree enable the RL \ChooseSubtree model to outperform the R-Tree more significantly.

\noindent \textbf{Query Processing Time.} Figure \ref{RL ChooseSubtree Query Processing Time} reports the query processing time of RL \ChooseSubtree on UNI dataset, where the left y-axis is the average relative query time and the right y-axis is the average query time. We observe that the average relative query time (green line) is generally consistent with the relative I/O cost of UNI reported in Figure \ref{RL insertion model performance}. Therefore, for the remaining of the paper, we report only relative I/O cost as a measure of query performance due to the space limit.

%
%



\begin{figure}[tbh]
\vspace*{-2mm}
\begin{center}
\noindent
  \includegraphics[width=0.85\linewidth]{Figures/RL_insertion_query_time_uni.pdf}
  \end{center}
  \vspace*{-5mm}
    \caption{RL \ChooseSubtree Query Processing Time}
    \label{RL ChooseSubtree Query Processing Time}
\end{figure}

\vspace{-1mm}
\subsubsection{RL Split.} \label{RL Splitting}
\begin{figure}[tbh]
\vspace*{-4mm}
\begin{center}
\noindent
  \includegraphics[width=0.85\linewidth]{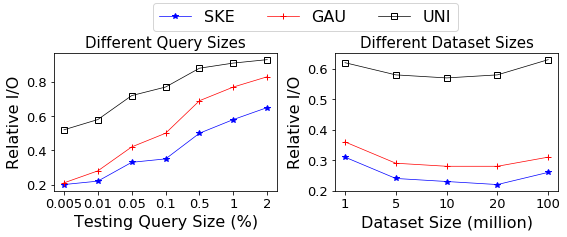}
  \end{center}
  \vspace*{-5mm}
    \caption{Performance of RL \Split}
    \label{RL splitting model performance}
\vspace*{-5mm}
\end{figure}

To evaluate the effect of RL \Split on query performance, Figure \ref{RL splitting model performance} report results on the 3 synthetic datasets of different sizes by running range queries of different sizes. 
RL \Split outperforms the R-Tree consistently over different query sizes, and the improvement can be up to 80\%.
We also observe that the RL \Split model has better improvement on the R-Tree as the query size decreases. Possible reasons would be similar as we discussed in Section \ref{RL Insertion Experiments} for RL \ChooseSubtree.





The RL \Split model also outperforms the R-Tree consistently over datasets of different sizes.
Note that  RL \Split  is trained on a small training dataset with only 100,000 data objects, but can be applied on large datasets successfully. 
Moreover, we also observe that 
as dataset size increases from 1 to 20 million, the improvement over the R-Tree generally increases on all 3 distributions.
However, model performance slightly deteriorates as dataset size increases from 20 million to 100 million. 
The reason might be due to the inherent complicatedness of \Split. Node splitting results in the creation of 2 new nodes. 
As dataset size increases, the increase in tree height further contributes to this complicatedness because more nodes may be split from one \Split process, and the RL \Split model trained on 100,000 data objects would have more room for improvement.

\vspace{-1mm}
\subsubsection{RLR-Tree.} \label{Combined RL Model}
This set of experiments is to evaluate the performance of RLR-Tree,
which is constructed from a combined RL \ChooseSubtree and RL \Split model.

\noindent \textbf{The Enhanced Training Process.} We first compare the performance of RL \ChooseSubtree, RL \Split, Naive RLR-Tree, which is obtained by directly applying RL \ChooseSubtree and RL \Split,
and RLR-Tree, which uses the enhanced training process (Section \ref{Combination of Insertion and Splitting Models}), on all the 5 datasets in terms of relative I/O cost. As shown in Table \ref{RL Insertion, RL Splitting and Combined RL Model Performance},
%
%
by applying the enhanced training process, RLR-Tree 
has the best performance on all datasets.




\begin{table}[tbh]
\normalsize
\centering
\vspace{-2mm}
\caption{RL \ChooseSubtree, RL \Split and RLR-Tree}
\vspace{-3ex}
\label{RL Insertion, RL Splitting and Combined RL Model Performance}
\begin{tabular}{| l | l | l | l | l | l |}
\hline
 & SKE & GAU & UNI & CHI & IND \\
\hline
RLR-Tree & \textbf{0.21} & \textbf{0.06} & \textbf{0.54} & \textbf{0.56} & \textbf{0.65} \\
\hline
Naive RLR-Tree & 0.24 & 0.08 & 0.58 & 0.61 & 0.67 \\
\hline
RL \ChooseSubtree & 0.29 & 0.08 & 0.56 & 0.60 & 0.67 \\
\hline
RL \Split & 0.22 & 0.28 & 0.58 & 0.63 & 0.71 \\
\hline
\end{tabular}
\end{table}

\begin{figure}[tbh]
\vspace*{-4mm}
\begin{center}
\noindent
  \includegraphics[width=0.85\linewidth]{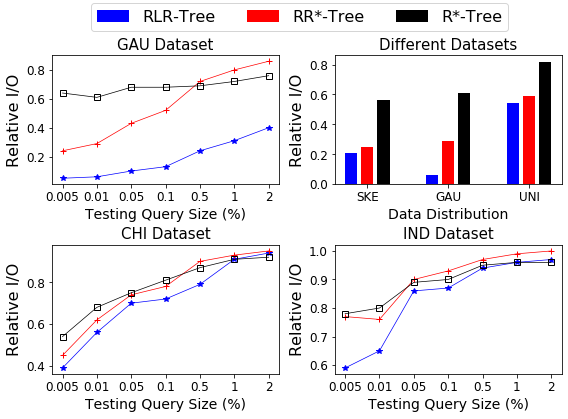}
  \end{center}
  \vspace*{-5mm}
    \caption{RLR-Tree Performance (Range Queries)}
    \label{The Combined RL Model Performance}
\vspace*{-4mm}
\end{figure}

\noindent \textbf{Range queries.} We evaluate the query performance of the RLR-Tree on all the 5 datasets using range queries of different sizes. Experimental results are shown in Figure \ref{The Combined RL Model Performance}. We observe that RLR-Tree outperforms the three baselines, the R-Tree, the R*-Tree, and the RR*-Tree, on all the synthetic datasets. The best performance is observed on the GAU dataset where RLR-Tree outperforms RR*-Tree by 78.6\% and R*-Tree by 92\% for query size 0.005\%. 
On the other hand, on the 2 real datasets, RLR-Tree outperforms R*-Tree, RR*-Tree and LISA by up to 27.3\%, 22.9\% and 40\% respectively. 
On all the 5 datasets, RLR-Tree has more significant advantages over baselines for smaller query sizes. 
This observation is consistent with RL \ChooseSubtree and RL \Split and possible reasons have been discussed in section \ref{RL Insertion Experiments}. 
For query size 2\% all the indices have similar performance, which is similar to the results reported in previous work~\cite{qi2018parallelizability}.



\begin{figure}[tbh]
\vspace*{-4mm}
\begin{center}
\noindent
  \includegraphics[width=0.85\linewidth]{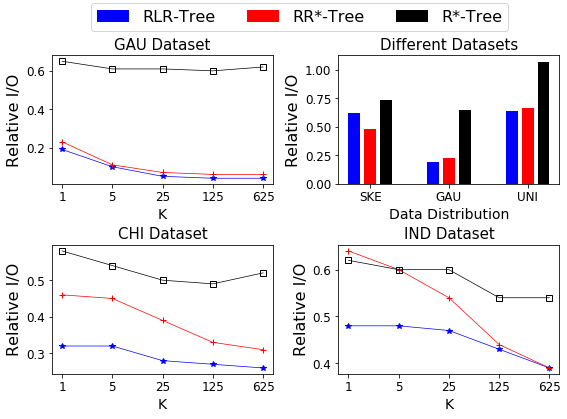}
  \end{center}
  \vspace*{-5mm}
    \caption{RLR-Tree Performance (KNN Queries)}
    \label{The Combined RL Model Performance KNN}
\vspace*{-4mm}
\end{figure}


\noindent \textbf{KNN queries.} To evaluate the performance of the RLR-Tree for types of queries that are not used in the model training process, we look into the K-Nearest-Neighbor (KNN) queries, which is a type of very popular spatial queries. A KNN query returns the $K$ nearest objects to a given query point. In our experiments, we consider different $K$ values, i.e. $K \in \{1,5,25,125,625\}$ with $K = 1$ being the default. For each $K$ value, 1,000 uniformly distributed query points are randomly generated in the data space. We use the algorithms proposed in \cite{roussopoulos1995knnquery} to compute KNN queries accurately. 
In Figure \ref{The Combined RL Model Performance KNN},
we observe that the RLR-Tree outperforms R*-Tree, RR*-Tree and LISA by up to 93.5\%, 30.4\% and 42\% respectively. 
The RLR-Tree outperforms all the baselines in almost all the cases except for the SKE dataset. 
%
%
We also observe that the relative query performance of RLR-Tree to R-Tree gets better for larger $K$ values. 
 %
%
The finding that the RLR-Tree also outperforms the baselines is particularly interesting —— the RLR-Tree is designed and trained to optimize the performance of range queries, rather than KNN queries. 
%
Additionally, we compute the reward and design the state features for the RLR-Tree in a way such that the model is trained to minimize the number of nodes accesses when answering range queries and not KNN queries.
%
However, despite those design features that do not favour KNN queries, RLR-Tree still has the best performance most of the time for KNN queries. 

\vspace{-2mm}
\subsubsection{Effect of Data Change} \label{dynamic updates}
We run a series of experiments to evaluate the robustness of the RLR-Tree when changes in data distribution are encountered in dynamic environments.

\begin{figure}[tbh]
\vspace*{-2mm}
\centering
\begin{subfigure}[b]{0.5\linewidth}
    \centering
    \includegraphics[width=\linewidth]{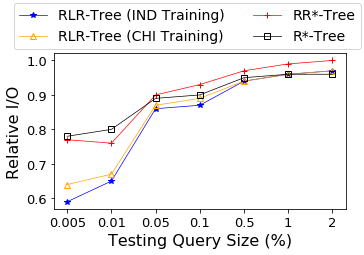}
    \vspace*{-6mm}
    \caption{IND Dataset}
    \label{Apply trained models on different data distributions IND Dataset}
\end{subfigure}%
\begin{subfigure}[b]{0.5\linewidth}
    \centering
    \includegraphics[width=\linewidth]{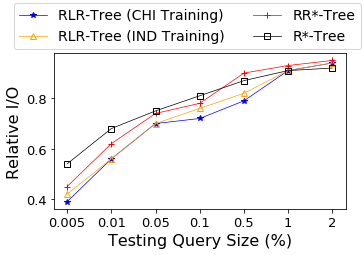}
    \vspace*{-6mm}
    \caption{CHI Dataset}
    \label{Apply trained models on different data distributions CHI Dataset}
\end{subfigure}
\begin{subfigure}[b]{0.5\linewidth}
    \centering
    \includegraphics[width=\linewidth]{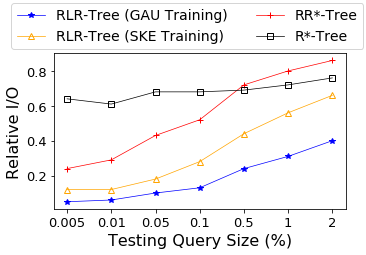}
    \vspace*{-6mm}
    \caption{GAU Dataset}
    \label{Apply trained models on different data distributions GAU Dataset}
\end{subfigure}%
\begin{subfigure}[b]{0.5\linewidth}
    \centering
    \includegraphics[width=\linewidth]{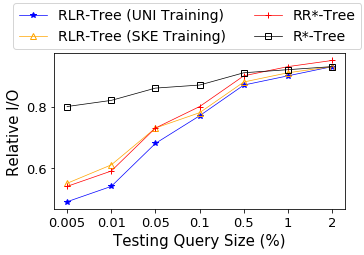}
    \vspace*{-6mm}
    \caption{UNI Dataset}
    \label{Apply trained models on different data distributions UNI Dataset}
\end{subfigure}
\vspace*{-8mm}
\caption{Apply Trained Models on Other Data Distributions}
\label{Apply trained models on different data distributions}
\vspace*{-3mm}
\end{figure}

\noindent {\bf Apply Trained Models on Different Data Distributions.} 
In this set of experiments, we train RL \ChooseSubtree and RL \Split models on one dataset and apply the trained models to build RLR-Tree on a different dataset. Figure \ref{Apply trained models on different data distributions} reports relative I/O cost of the evaluated methods. 
For example, in Figure \ref{Apply trained models on different data distributions IND Dataset}, we use the models trained on CHI and IND, respectively, to build different RLR-Trees on IND.  
On 
CHI and IND, we observe that though the models are trained on a different dataset, the constructed RLR-Trees still outperform all baselines for all query sizes except 2\%. Note that compared with the RLR-Trees that are constructed by models trained on the same dataset, these RLR-Trees only experience a small degree of performance deterioration. This is perhaps because these real datasets share  common features as they mostly consist of "clusters" of high data density which represent developed regions surrounded by vast regions of low data density which represent rural areas. In contrast, for synthetic datasets, we use the models trained on the SKE dataset to build RLR-Trees for the GAU dataset and the UNI dataset. We observe that the performance deterioration of the RLR-Trees constructed with models trained on a different dataset is more significant. On GAU, although the RLR-Tree with SKE training is able to outperform all baselines for all query sizes, its query performance is up to 57\% poorer than the RLR-Tree with GAU training. On UNI, we observe that the  performance of RLR-Tree with SKE training is generally close to that of the RR*-Tree.

\begin{figure}[tbh]
\vspace*{-4mm}
\begin{center}
\noindent
  \includegraphics[width=0.85\linewidth]{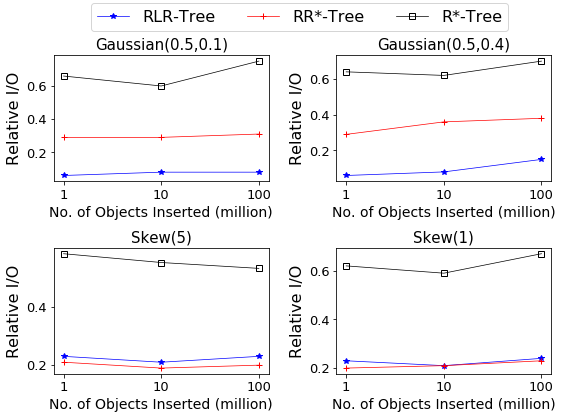}
  \end{center}
  \vspace*{-4.5mm}
    \caption{Updates with Different Data Distributions}
    \label{Dynamic Updates with Different Data Distributions}
\end{figure}

\noindent {\bf Dynamic Updates with Gradual Data Change.} 
%
This experiment is to investigate how well the RLR-Tree handles dynamic updates with gradual change of data distribution. Specifically, we first train and build 2 RLR-Trees of size 1 million using the GAU dataset with ($\mu = 0.5$, $\sigma = 0.2$) and the SKE dataset with ($c=9$) with their default settings respectively. For the RLR-Tree built using the GAU dataset, we insert up to 100 million spatial objects from GAU distribution with $(\mu = 0.5$, $\sigma = 0.1)$ and $(\mu = 0.5$, $\sigma = 0.4)$, and report the average relative I/O cost of 1,000 random queries, respectively. For the RLR-Tree built using the SKE dataset, we insert up to 100 million spatial objects from SKE distribution with $(c = 5)$ and $(c = 1)$, and report the average relative I/O cost of 1,000 random queries, respectively. Relevant experimental results are shown in Figure \ref{Dynamic Updates with Different Data Distributions}. We observe that despite inserting up to 100 times as many data objects from a different data distribution, RLR-Tree generally does not experience obvious performance deterioration. For the GAU dataset, RLR-Tree consistently outperforms all baselines. On the other hand, for the SKE dataset, RLR-Tree outperforms R*-Tree significantly but is slightly outperformed by RR*-Tree.


\vspace{-2mm}
\subsubsection{The Effects of Parameters} \label{subsed:exp:parameter}
\hfill

\begin{figure}[tbh]
\vspace*{-4mm}
\begin{center}
\noindent
  \includegraphics[width=0.9\linewidth]{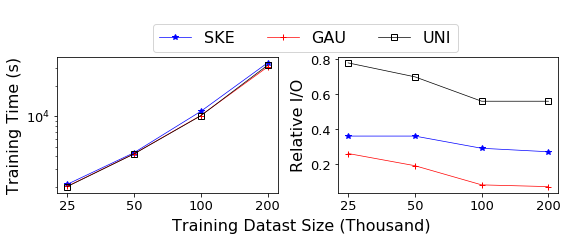}
  \end{center}
  \vspace*{-5mm}
    \caption{Effect of Varying Training Dataset Sizes}
    \label{Effect of Varying Training Dataset Sizes}
\end{figure}

\noindent{\bf Training Dataset Size.} 
\noindent 
This experiment is to evaluate the effect of training dataset size on the performance of the RLR-Tree. 
We would expect better query performance if we train the RL \ChooseSubtree and RL \Split models on the full dataset. However, the training is  slow on large datasets. Instead, we propose to use a small training dataset. Experimental results are shown in Figure \ref{Effect of Varying Training Dataset Sizes}.
%
%
%
%
%
First, we observe that the training time of RL \ChooseSubtree model  for different data distributions is similar, and increases significantly with the size of the training dataset. 
As expected, the query performance of the trained models improves as training dataset size increases from 25,000 to 100,000. However, the query performance becomes stable after the dataset size reaches 100,000, and the improvement of using training dataset of size 200,000 over 100,000 is not significant. 
%
%
%
The results for RL Split are qualitatively similar to those for RL \ChooseSubtree. 
Therefore we set training dataset size to be 100,000 by default which achieves a good tradeoff between training time and query performance.
%
Note that 
RLR-Tree only needs to be trained once on a small training dataset and the learned models can be used on large dataset to build index and then to handle updates.


\begin{figure}[tbh]
\vspace*{-4mm}
\centering
\begin{subfigure}[b]{0.49\linewidth}
    \centering
    \includegraphics[width=\linewidth]{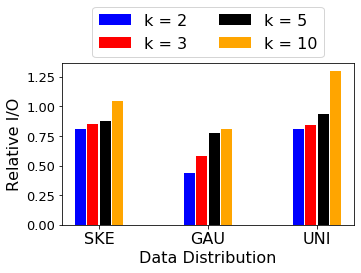}
    \vspace*{-6mm}
    \caption{Varying $k$ Values}
    \label{Effect of Varying $k$ Values}
\vspace*{-4mm}
\end{subfigure}%
\begin{subfigure}[b]{0.49\linewidth}
    \centering
    \includegraphics[width=\linewidth]{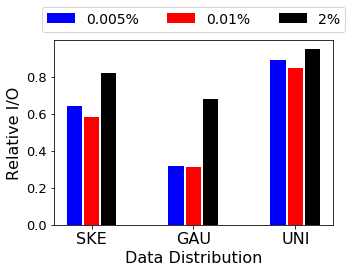}
    \vspace*{-6mm}
    \caption{Varying Training Query Sizes}
    \label{Effect of Varying Training Query Sizes}
\vspace*{-4mm}    
\end{subfigure}
\caption{The Effects of Parameters}
\label{The Effects of Parameters}
\vspace*{-4mm}
\end{figure}

\noindent {\bf The Value of $k$.} 
As shown in Figure \ref{Effect of Varying $k$ Values}, as top $k$ candidates are shortlisted to form the action space, the value of $k$ has a direct impact on the query performance of the resulted RLR-Tree. 
On one hand, when $k$ is larger, more actions are available to be selected for the trained model.
On the other hand, model performance can be adversely affected when the action space is large as the trained model may not do a good job to filter out "bad" candidates. To find a good $k$ value, we test different values of $k$ on the 3 synthetic datasets of size 500,000. 

We use RL \ChooseSubtree as an example to show the effect of $k$.
We observe qualitatively similar trends on all the 3 synthetic datasets.
%
We make 2 observations. Firstly, recall that $k=1$ is the trivial case that generates the reference tree. By including one additional candidate in the action space, i.e., when $k=2$, we achieve significant query performance improvement for RL \ChooseSubtree. The best performance improvement is $56\%$ on the GAU dataset. 
Secondly, we observe that RL \ChooseSubtree has the best result at $k=2$ for all the 3  datasets. As $k$ value increases, model performance deteriorates gradually. This observation aligns with our expectation. When $k$ value approaches and exceeds 10, the RL \ChooseSubtree model starts to fail to outperform the R-Tree.
%
We observe similar trends for RL \Split on the 3 datasets, and do not report the result here due to space limitation.




\noindent {\bf Training Query Size.} 
%
%
%
This experiment is to evaluate the effect of the training query size on the query performance of the resulted RLR-Tree. 
Figure \ref{Effect of Varying Training Query Sizes}
reports the average query performance of RL \ChooseSubtree  for each training query size for each dataset.
We observe that the query performance of RL \ChooseSubtree is rather poor when using the largest training query size, i.e. 2\%. On the GAU dataset, the model performance with training query size of 2\% is more than 100\% poorer than that with our default setting of 0.01\%. On the other hand, when using training query size of 0.005\%, model performance is on par with our default setting on GAU dataset and shows slightly poorer results on SKE and UNI datasets. Therefore, we set the training query size to be 0.01\% by default.
We observe similar trends for RL \Split.

\vspace{-1mm}
\subsubsection{Other Experiments} \label{subsec:other exp} \hfill

\begin{figure}[tbh]
\centering
\begin{minipage}{0.49\linewidth}
    \centering
    \includegraphics[width=\linewidth]{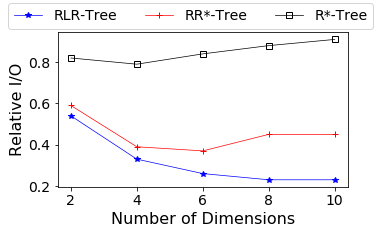}
    \vspace{-8mm}
    \caption{Changing Dimensions}
    \label{UNI Dataset with Changing Dimensions}
\vspace{-3mm}
\end{minipage}
\begin{minipage}{0.49\linewidth}
    \centering
    \includegraphics[width=\linewidth]{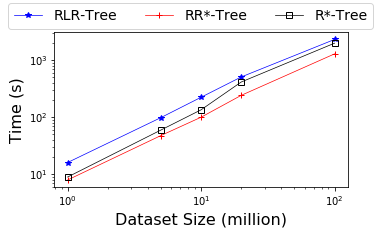}
    \vspace{-8mm}
    \caption{Index construction time}
    \label{Index construction time}
\vspace{-3mm}  
\end{minipage}
\end{figure}

\noindent \textbf{Number of Dimensions.}
This experiment is to show how well RLR-Tree scales with dimensions. We vary the number of dimensions from 2 to 10. For each case, we generate a synthetic dataset with 20 million objects following the Uniform distribution. The Uniform distribution is chosen because it is commonly used for high dimension synthetic datasets in previous works \cite{beckmann2009revisedrstartree, nathan2020flood}. For each dataset, we report the average relative I/O cost for answering 1,000 random queries for each index in Figure \ref{UNI Dataset with Changing Dimensions}. The average data selectivity of the query workload for each dataset is kept constant at 0.01\%. We observe that RLR-Tree consistently outperforms all baselines all the time. Moreover, its advantage becomes more significant as the number of dimensions increases, which illustrates the robustness of RLR-Tree w.r.t. the number of dimensions. 

\noindent \textbf{Index Construction Time and Size.} 
The index construction time is similar for different data distributions of the same size, and we use the GAU dataset as an example. As shown in Figure \ref{Index construction time}, index construction time increases almost linearly with increase in dataset size. 
Note that the construction time of RLR-Tree is comparable to that of R*-Tree and RR*-Tree. 
And we expect RLR-Tree construction time to be shorter when using better GPU devices.
For index size, the RLR-Tree and other baselines almost have the same size. Therefore, we report the RLR-Tree sizes for datasets of different sizes. As shown in Table \ref{Index size for Uniform datasets}, RLR-Tree size increases linearly as dataset size increases.


\begin{table}[tbh]
\normalsize
\centering
\vspace{-3mm}
\caption{Index size for GAU datasets}
\vspace{-3ex}
\label{Index size for Uniform datasets}
\begin{tabular}{| l | l | l | l | l | l |}
\hline
\textbf{Dataset size (million)} & 1 & 5 & 10 & 20 & 100 \\
\hline
\textbf{RLR-Tree size (MB)} & 39 & 195 & 390 & 780 & 3900 \\
\hline
\end{tabular}
\vspace*{-4mm}
\end{table}

\section{Related Work} \label{Related Work}


\subsection{Spatial Indices} \label{Spatial Data Indexes}

We next introduce three categories of spatial indices.

\noindent \textbf{Data partitioning based indices: R-Tree and its variants.} 
The category of indices partitions the dataset into 
subsets and indexes the subsets.
Typical examples include the R-Tree and its variants,
%
such as
R$^*$-Tree \cite{beckmann1990rstar},
R$^+$-Tree \cite{sellis1987rplus}, RR$^*$-Tree \cite{beckmann2009revisedrstartree}. 
An R-Tree is usually built through a one-by-one insertion approach, and is maintained by dynamic updates. 
%
%
%
%
As discussed in Section~\ref{Introduction}, these R-Tree variants are equipped with 
various heuristic rules for the two key operations of building an R-Tree, \ChooseSubtree or \Split, but none of the rules have dominant query performance. 
This partly motivates us to design learning-based solutions for \ChooseSubtree and \Split operations in this work.

There exist R-Tree variants that explore the query workload property or the special property of data. For example, in QUASII \cite{pavlovic2018quasii} and CUR-Tree  \cite{ross2001curtree}, 
the query workload is utilized to build an R-Tree. 
%
These extensions are orthogonal to our work. 


\noindent \textbf{Data partitioning based indices: Packing.} An alternative way of building an R-Tree is to pack data points into leaf nodes and then build an R-Tree bottom up.
The original R-Tree and almost all of its variants are designed for a dynamic environment,
being able to handle insertions and deletions, while packing methods are for static databases \cite{kamel1993packing}.  
Packing needs to have all the data before building indices, which are not always available in many situations.
%
%
The existing packing methods explore different ways of sorting spatial objects to achieve better ordering of the objects and eventually better packing. Some ordering methods \cite{arge2008priorityrtree} are based on the coordinates of objects, such as STR~\cite{leutenegger1997str}, TGS~\cite{garcia1998tgs} and the lowx packed R-Tree \cite{roussopoulos1985lowxpack} and other ordering methods are based on space filling curves such as z-ordering \cite{orenstein1986zorder,qi2018parallelizability}, Gray coding \cite{faloutsos1986graycode} and the Hilbert curve \cite{faloutsos1989hilbertcurve}. 
 



\noindent \textbf{Other spatial indices.}
In space partitioning based indices, such as kd-Tree \cite{bentley1975kdtree} and Quad-Tree \cite{finkel1974quadtree}, the space is recursively partitioned until the number of objects in a partition reaches a threshold. 
%






\subsection{Learned Indices} \label{Machine Learning and Data Structures}



\noindent \textbf{Learned one-dimensional indices.} 
The idea behind learned index is to learn a function that maps a search key to the storage address of a data object. The idea of learned indices is first introduced by \cite{kraska2018case}, which proposes the Recursive Model Index (RMI)  to learn the data distribution in a hierarchical manner. 
It essentially learns a cumulative distribution function (CDF) using a neural network to predict the rank of a search key. 
%
%
The idea inspires a number of follow-up studies~\cite{ding2020alexindex, ferragina2020pgmindex, kipf2020radixsplineindex, wu2021updatable} on learned indices. 

%


\noindent \textbf{Learned spatial indices.} 
Inspired by the idea of learned one-dimensional indices, several learned spatial indices have been proposed. 
%
%
%
The Z-order model~\cite{wang2019learnedrtree} extends RMI to spatial data by using a 
 space filling curve to order data points and then learning the CDF to map the key of a data point to its rank.
%
Recursive spatial model index (RSMI)~\cite{qi2020deeplearningrtree} further develops the Z-order idea \cite{wang2019learnedrtree} and RMI. 
RSMI maps the data points to a rank space using the rank space-based transformation technique~\cite{qi2018parallelizability}. 
%
%
%
%
LISA \cite{li2020lisa} is a disk-based learned index.
It partitions the data space with a grid,  numbers the grid cells, and learns a data distribution based on this numbering. 
%
Similar to Z-order model~\cite{wang2019learnedrtree}  and RSMI~\cite{qi2020deeplearningrtree}, 
Flood \cite{nathan2020flood} 
also maps a dataset to a uniform rank space before learning a CDF. Differently, it utilizes workload to optimize the learning of the CDF, and it learns the CDF of each dimension separately. 
%
Tsunami \cite{ding2020tsunami} extends Flood to better utilize workload to overcome the limitation of Flood in handling skewed workload and correlated data.  
 The ML-Index \cite{davitkova2020mlindex} generalizes the idea of the iDistance scaling method \cite{jagadish2005idistance} to map point objects to a one-dimensional space and then learns the CDF.


\noindent {\bf Remark.}
These learned indices all aim to learn a CDF for a particular data to replace the traditional indices. However, RLR-Tree is fundamentally different ——
Instead of learning any CDF, we train RL models to handle \ChooseSubtree and \Split operations.
Furthermore, these learned indices have the following limitations compared with our solution: 
First, they can only handle spatial point objects while our proposed method is able to handle any spatial data, such as rectangular objects.
Second, they all need customized algorithms to handle each type of query. They focus on certain types of queries and it is not clear how they can process other types of queries. For example, some of them  ~\cite{nathan2020flood, ding2020tsunami, davitkova2020mlindex} do not consider KNN queries, an important type of spatial queries; 
Some learned indices \cite{li2020lisa} extend their algorithm for range queries to handle KNN queries by issuing a series of range queries until $k$ points are found. However, the query performance largely depends on the size of the region used. 
In contrast, the RLR-Tree simply uses existing query processing algorithms for R-Tree to handle different types of queries.
Third, some of these learned indices \cite{wang2019learnedrtree, qi2020deeplearningrtree} return approximate query results while our query results are accurate.  
Fourth, both Flood and Tsunami need the query workload as the input. However, we do not assume the query workload to be known. Finally, updates are not discussed for Flood, Tsunami or the ML-Index. Although RSMI~\cite{qi2020deeplearningrtree} and LISA~\cite{li2020lisa} can handle updates, their models have to be retrained periodically to retain good query performances. In contrast, our proposed RLR-Tree readily handles updates without the need to keep retraining the models.

\subsection{Applications of Reinforcement Learning} \label{Reinforcement Learning and Relevant Applications}

To the best of our knowledge, no work is done to use RL to improve on the R-Tree index or its variants. 
RL has been successfully applied to solve other database applications, such as database tuning tasks \cite{trummer2019skinnerdb, zhang2019endtoend, liang2019opportunistic}, 
similarity search \cite{wang2020efficient}, join order selection \cite{yu2020rltreelstm}, index selection \cite{sadri2020onlineindexselection}, and QD-Tree for data partitioning \cite{yang2020qdtree}. In particular, to build a QD-Tree, 
an RL model is trained to learn a policy to make partitioning decisions to maximize the data skipping ratio for a given query workload. 
The MDP design of our RLR-Tree is fundamentally different from that of QD-Tree, which is based on NeuroCuts \cite{liang2019neuralcuts}, in at least three aspects. 
1) State design: QD-Tree follows a tree-structured MDP where a cut at a node (state) creates 2 new nodes (next states). In contrast, in an RLR-Tree, an RL ChooseSubtree decision at a node (state) leads to the selected child node (next state) and an RL Split decision at a node (state) leads to the parent node (next state). 
2) Action space: For QD-Tree, the query workload is assumed to be known and candidate cuts are generated from a standard SQL planner. For RLR-Tree, we do not assume any prior knowledge of the query workload and have to come out with suitable actions ourselves.
3) Reward: Among others, QD-Tree computes rewards after the whole tree is built, while RLR-Tree computes rewards after each splitting/insertion process.
We would also like to highlight that QD-Tree cannot handle updates as the RLR-Tree does.
%
%

\section{Conclusions and Future Work} \label{Conclusion}

We propose the RLR-Tree to improve on the R-Tree and its variants designed for a dynamic environment.
Experimental results show that RLR-Tree is able to outperform  the RR*-Tree, the best R-Tree variant, by up to 78.6\% for range queries and up to 30.4\% for KNN queries. 
Although the models of the RLR-Tree are trained on a small training dataset of size 100,000, the trained models are readily applied on large datasets and are able to handle dynamic updates even with changes in data distribution.

This work take a first step to use machine learning to improve on the R-Tree, and we believe it we would open a few promising directions for future work: 1) further explore and refine the designs of the states, the action space and the reward signal; 2) extend the idea to index enriched spatial data, such as spatial-temporal data, moving objects, and spatial-textual data; 3) explore other machine learning models.


\normalem
\bibliographystyle{ACM-Reference-Format}
\bibliography{sample}

\end{document}

%% file: Introduction.tex
\vspace{-3mm}
\section{Introduction} \label{Introduction}


To support efficient processing of spatial queries, such as range queries and KNN queries, 
spatial  databases 
have 
relied on delicate indices. The R-Tree~\cite{guttman1984rtree}  is arguably the
most popular spatial index that prunes irrelevant data for
queries. 
%
%
R-Trees have attracted
extensive research interests \cite{beckmann1990rstar, sellis1987rplus, beckmann2009revisedrstartree, pavlovic2018quasii, ross2001curtree, kanth1997promo_demo_rtree, kamel1993packing, arge2008priorityrtree, leutenegger1997str, garcia1998tgs, roussopoulos1985lowxpack, qi2018parallelizability, wang2019learnedrtree, qi2020deeplearningrtree, li2020lisa, nathan2020flood, ding2020tsunami, davitkova2020mlindex, yang2020qdtree} 
and are widely used in commercial databases such as PostgreSQL and MySQL. 

The learned index has been proposed in \cite{kraska2018case}, which proposes a recursive model index (RMI) for indexing 1-dimensional data by learning a cumulative distribution function (CDF) to map a search key to a rank in a list of ordered data objects. To address the limitations of the RMI, such as the lack of supporting updates, and to improve it, several learned indices have been proposed based on the RMI. 
The idea of learned indices is also extended for spatial data~\cite{qi2020deeplearningrtree, wang2019learnedrtree,li2020lisa, pandey2020caseforspatialindex} and multi-dimensional data~\cite{ nathan2020flood, ding2020tsunami, davitkova2020mlindex}. 
They usually map spatial data points in a dataset to a uniform rank space (e.g., using a space filling curve), and then learn the CDF for this dataset. Despite the success of these learned indices in improving the efficiency of processing some types of queries, they still have various limitations, e.g., they can only handle spatial point objects and limited types of spatial queries, some only return approximate query results, and they either cannot handle updates or need a periodic rebuild to retain high query efficiency (Details in Section \ref{Related Work}). 
These limitations, together with the requirement that the learned indices need a replacement of the index structures and query processing algorithms currently used by spatial database systems, would make them not easy to be deployed in current database systems. 

In this work, rather than learning a CDF for spatial data, we consider a fundamentally different approach, i.e., to use machine learning techniques to construct an R-Tree in a data-driven way for better query efficiency in a dynamic environment where updates occur frequently and bulk loading is not viable. Specifically, we propose to build machine learning models for the two key operations of building an R-Tree, namely \ChooseSubtree and \Split operations, which currently rely on hand-crafted heuristic rules.
\emph{ Note that we do not modify the basic structure of the R-Tree and thus all the currently deployed query processing algorithms will still be applicable to our proposed index. }
This would make it easier for the learning based index to be deployed by current databases. 

To motivate our idea, we next revisit \ChooseSubtree and \Split. When inserting a new spatial object, the \ChooseSubtree operation is invoked iteratively, i.e., choosing which child node to insert the new data object, until a leaf node is reached. 
If the number of entries in a node exceeds the capacity, the Split operation is invoked to divide the entries into two groups.
%
%
Many R-Tree variants have been proposed, which mainly differ in their strategies for the insertion of new objects (i.e., \ChooseSubtree) as well as algorithms for splitting a node (i.e., \Split). 
Almost all these strategies are based on hand-crafted heuristics. However, there is no single heuristic strategy that dominates the others in terms of  query performance. To illustrate this, we generate a dataset with 1 million uniformly distributed data points and construct four R-Tree variants using four different Split strategies, namely \emph{linear} \cite{guttman1984rtree}, \emph{quadratic} \cite{guttman1984rtree}, \emph{Greene's} \cite{greene1989greenesplitting} and \emph{R*-Tree} \cite{beckmann1990rstar}. 
%
We run 1,000 random range queries and rank the four indices based on the query processing time of each individual query. 
We observe that no single index has the best performance for all the queries. For example, Greene's Split has the best query performance among 50\% of the queries while R*-Tree Split is the best for 49\% of the queries.
The observation that no single index dominates the others holds on other datasets as well, and the top performers may be different on different datasets.

The observation motivates us to develop machine learning models to handle the \ChooseSubtree and \Split operations, to replace the  heuristic strategies used in the R-Tree and its variants.
Furthermore, we observe that the two operations can be considered as two sequential decision making problems. Therefore, we model them as two Markov decision processes (MDPs) \cite{puterman2014markov} and propose to use reinforcement learning (RL) to learn models for the two operations.
%
%

However, it is very challenging to make the idea work---We have tried various ways to define the two MDPs, which significantly affect the performance of the learned models. 
The first challenge 
is how to formulate \ChooseSubtree and \Split as MDPs. How should we define the states, actions, and 
the reward signals for each MDP? Specifically, 
1)~\emph{Designing the action space}. A straightforward idea could be to define an action as one of the existing heuristic strategies. 
However, the idea did not work, for which we observed from the experiments that different strategies would often make the same insertion/splitting decision, and thus it leaves us very little room for improvement. 
Some other possible ideas include defining larger action spaces, but then it would increase the difficulty of training the model.
%
2)~\emph{Designing the state}. As the number of entries varies across different R-Tree nodes, it is challenging to represent the state of a node for both operations. 
3)~~\emph{Designing the  reward signal}. It is nontrivial to design 
a function that evaluates the reward of past actions during model training to encourage the RL agent to take ``good'' actions in our problem.

The second challenge is how to use RL to address the defined MDPs. For instance, in the construction of an R-Tree, node overflow (and thus the \Split operation) occurs less frequently than the \ChooseSubtree operation. 
Therefore, only a few state transitions for Split operations are generated, making it difficult for the RL agent to learn useful information. Moreover, a previous decision (\ChooseSubtree or \Split) may affect the structure of the R-Tree in the future. Therefore, a ``good'' action may receive a bad reward due to some bad actions that were made previously. 
This makes it even more challenging to learn a good policy to solve the two MDPs.

\noindent \textbf{Our method.} 
To this end, we propose the RL based R-Tree, called RLR-Tree. In the RLR-Tree, we carefully model the two MDPs, \ChooseSubtree and \Split, and propose a novel RL-based method to learn policies to solve them. The learned policies replace the hand-crafted heuristic rules to build a different R-Tree, namely RLR-Tree.
The RLR-Tree possesses several salient features: 

(1) Its models are trained on a small training dataset and then used to 
build an RLR-Tree on datasets of much larger scales. 
Furthermore, models trained on one data distribution can be applied to data with a different distribution to build an RLR-Tree, the RLR-Tree still significantly outperforms the R-Tree and its variants in answering spatial queries.

(2) It achieves up to 95\% better query performance than the R-Tree and its variants that are designed for a dynamic environment. 

(3) Like the R-Tree, the RLR-Tree can be built on spatial objects of different types, such as points or rectangles. However, to the best of our knowledge, existing learned indices \cite{qi2020deeplearningrtree, nathan2020flood, ding2020tsunami, davitkova2020mlindex, li2020lisa, wang2019learnedrtree} can only handle point objects when used for spatial data.

(4) The RLR-Tree simply deploys any existing query processing algorithms for R-Trees to answer queries of different types, without the need of designing new query processing algorithms. However,  other learning based indices \cite{qi2020deeplearningrtree, nathan2020flood, ding2020tsunami, davitkova2020mlindex, li2020lisa, wang2019learnedrtree} often only focus on limited types of queries, e.g., range queries, and they need to design new algorithms for each type of query.

%


%

In summary, we make the following contributions:
\vspace*{-1ex}

\smallskip
    (1) We propose to 
    train machine learning models to replace heuristic rules in the construction of an R-Tree to improve on its query efficiency in a dynamic environment where updates occur frequently and bulk loading is not viable. To the best of our knowledge, this is the first work that uses machine learning to improve on the R-Tree without modifying its structure; Therefore, all currently deployed query processing algorithms are still applicable and the proposed index can be easily deployed by current databases.
    
    (2) We model the \ChooseSubtree and the \Split operations as two MDPs, and carefully design their states, actions and reward signals. We also present some of our unsuccessful trials of designing.
    
    (3) We design an effective and efficient learning process that learns good policies to solve the MDPs. The learning process enables us to apply our RL models trained with a small dataset to build an R-Tree for up to more than 100 million spatial objects.
    
    (4) We conduct extensive experiments on both real and synthetic datasets. The experimental results show that our proposed index achieves up to 95\% better query performance for range queries and 96\% for KNN queries than R-Tree and its variants, and up to 40\% better query performance for range queries and 42\% for KNN queries than LISA~\cite{li2020lisa}, which is the only disk based learned spatial index that returns exact results for range queries and KNN queries.

%% file: problem.tex
\section{Preliminary and Problem} \label{Learning to Build R-Tree}

\subsection{Preliminary} \label{Preliminary}
R-Tree \cite{guttman1984rtree} is a balanced tree for indexing multi-dimensional objects, such as coordinates and rectangles. 
Each tree node can contain at most $M$ entries. Each node (except the root node) must also contain at least $m$ entries. Each entry in a non-leaf node consists of a reference to a child node and the minimum bounding rectangle (MBR) of all entries within this child node. Each leaf node contains entries, each of which consists of a reference to an object and the MBR of that object. Therefore, a query that does not intersect with the MBR cannot intersect any of the contained objects. 

The algorithms for building an R-Tree comprise two key operations, \textbf{ChooseSubtree} and \textbf{Split}.
To insert an object into an R-Tree, 
starting from the root node, \ChooseSubtree is iteratively invoked to decide in which subtree to insert the object, until a leaf node is reached. The object is inserted into the leaf node and its corresponding MBR is updated accordingly. 
If the number of entries in a leaf node exceeds $M$, the \Split operation is invoked to divide the objects into two groups: one remains in the original leaf node and the other will become a new leaf node. The Split operation may be propagated upwards as an entry referring to the new leaf node is added to its parent node, which may overflow and need to be split. 

The query performance of an R-Tree highly depends on how the R-Tree is built. 
Many R-Tree variants have been proposed with different \ChooseSubtree and \Split strategies as discussed in Section~\ref{Introduction}.

\vspace*{-2mm}
\subsection{Problem Statement}
As discussed in Section~\ref{Introduction}, most of the existing R-Tree variants adopt hand-crafted \ChooseSubtree and \Split strategies, and no strategy can build an R-Tree with dominant query performance in all cases. Motivated by this, we aim to learn to build an R-Tree, 
i.e., using RL models to make decisions for \ChooseSubtree and \Split instead of relying on heuristic rules.
The new index is called RLR-Tree.



%% file: method.tex
\section{RLR-Tree} \label{RLR-Tree}

\subsection{Overview}
The process of inserting a new object into an R-Tree is essentially a 
combination of two typical sequential decision making processes. In particular, 
starting from the root, it needs to make a decision on  which child node to 
insert the new object at each level in a top-down traversal (\ChooseSubtree). It also needs to  
make a decision on how to split an overflowing node and divide the entries in a
bottom-up traversal (\Split). 
Reinforcement learning (RL) has been proven to be effective in 
solving sequential decision making problems. Therefore, we propose to 
model the insertion of a new object as a combination of two Markov decision 
processes (MDPs) and adopt RL to learn the optimal policies 
for \ChooseSubtree and \Split operations.

Figure \ref{RLR-Tree Overview} depicts an overview of the proposed solution to build an RLR-Tree, as well as using the RLR-Tree to answer queries.
In offline training, we propose new solutions to train RL \ChooseSubtree and RL \Split models using a small dataset or a subset.
%
This is the focus of this work.
Note that 
models trained on one dataset is readily applied on a different dataset as shown in experiments.
The two trained models can be integrated into the algorithms for R-Tree construction to build the RLR-Tree and R-tree maintenance with dynamic updatets.
Finally, any existing query processing algorithm designed for the R-Tree family can be used for RLR-Tree to answer different types of spatial queries. 

\begin{figure}[tbh]
\vspace*{-3mm}
\begin{center}
\noindent
  \includegraphics[width=0.95\linewidth]{Figures/RLR_tree_overview.pdf}
  \end{center}
  \vspace*{-4mm}
    \caption{RLR-Tree Overview}
    \label{RLR-Tree Overview}
\vspace*{-3mm}
\end{figure}

%


Next, we focus on the offline training and present our final designs for the two MDPs and the other representative designs that we have explored to formulate the problem. We present RL \ChooseSubtree and its model training in 
Section~\ref{sec:subtree_selection}, and RL \Split and its model training in Section~\ref{sec:node_splitting}. We present how to train the 
two models together in Section~\ref{Combination of Insertion and Splitting Models}. 
We briefly introduce how to integrate the trained models into existing algorithms to construct the RLR-Tree and then to handle updates in Section \ref{construction}.
%


\vspace*{-3mm}
\subsection{ChooseSubtree}\label{sec:subtree_selection}
To insert a new object into the R-Tree, we need to conduct a 
top-down traversal starting from the root. In each node, we need to decide which
child node to insert the new object.
To choose a subtree with RL, we formulate this problem as an MDP. We proceed to
present how to train a model to learn a policy for the MDP.

\vspace*{-2mm}
\subsubsection{MDP Formulation}
An MDP consists of four components, namely \emph{states}, \emph{actions}, 
\emph{transitions}, and \emph{rewards}. We proceed to explain how states 
and actions are represented in our model, and then present the reward signal design, which is particularly challenging. 

\vspace*{-1mm}
\smallskip
\noindent\textbf{MDP: State Space.}
A state captures the environment that is taken into account for decision making. 
For \ChooseSubtree, it is a natural idea that a state is from the tree node 
whose child nodes are to be selected for inserting a new object. The challenging 
question is: what kind of information should we extract from the 
tree node to represent the state?

Intuitively, as we need to decide which child node to insert the new object, 
it is necessary to incorporate the change of the child node if we add the new 
object into it for each child node. Possible features that capture the change 
of a child node $N$ include: (1) $\Delta Area(N, o)$, which is the area 
increase of the MBR of $N$ if we add the new object $o$ into $N$; (2) $\Delta
Peri(N, o)$, which is the perimeter increase of the MBR of $N$ if we add $o$ 
into $N$, and (3) $\Delta Ovlp(N, o)$, which is the increase of the overlap 
between $N$ and other child nodes after $o$ is inserted into $N$. Furthermore, 
it is helpful to know the occupancy rate of the child node, denoted by $OR(N)$, 
which is the ratio of the number of entries to the capacity. A child node with a 
high occupancy rate is more likely to overflow in the future.

As we have presented several features to capture the properties of a tree node,
a straightforward idea is that we compute the aforementioned features for every
child node and concatenate them to represent the state. However, 
the number of child nodes varies across different nodes, making it 
difficult to represent a state with a vector of a fixed length. An  idea 
to address this challenge is to do padding, i.e., to append zeros to the 
features of the child nodes to get a $4\cdotp M$ dimensional vector, as there are
four features and there are at most $M$ child nodes. However, 
the padded representations are likely to have 
many zeros which will add noises and mislead the model, resulting in 
poor performance. This is confirmed by our preliminary experiments. 

To address the challenge, we propose to only use a small part of child nodes to define the state.
This is because most of the child nodes are not good candidates for hosting 
the new object, as inserting the new object  may greatly increase their MBRs.
Here we aim to 
prune unpromising child nodes from the state space, and  our RL agent will not consider them for 
representing a state. Our design of state representation is as follows: We first 
retrieve the top-$k$ child nodes in ascending order of area increase, where the choice of area increase is based on empirical findings.
Then for each retrieved child node, we compute four features $\Delta Area(N, o)$,
$\Delta Peri(N, o)$, $\Delta Ovlp(N, o)$, and $OR(N)$. We concatenate the 
features of the $k$ child nodes to get a $4\cdotp k$ dimensional vector to 
represent a state. 
$k$ is a parameter to be set empirically. 
Note that to make the representation of different states comparable, the 
increases of area, perimeter and overlap are normalized by the maximum 
corresponding value among all $k$ child nodes.

\noindent\textbf{Remark.}
It is a natural idea that we can include more features to represent a 
state. For instance, we can include global information, such as the tree depth 
and the size of the tree, and local information, such as the depth of the tree 
node, the coordinates of the boundary of the MBR. However, our experiments show
that these features do not improve the performance of our model while making the model training slower. The four 
features that we use are sufficient to train our model to make good subtree 
choices as shown in our experiments. 
It would be a useful future direction to design and evaluate other state features.

\noindent\textbf{MDP: Action Space.}
As many 
R-Tree variants have been proposed with different \ChooseSubtree strategies, 
such as minimizing the increase of area, perimeter, or overlap. A 
straightforward idea 
is to make the different cost 
functions the actions, i.e., to decide which cost function to use 
to choose the subtree. 
After trying different combinations of these cost functions and different state space designs, this idea is proven to be ineffective by our experimental results. 
Table~\ref{Cost Function Based Action Space vs Child Node Based Action Space} 
depicts the average relative I/O cost for processing 1,000 random range queries on three datasets of different distributions, namely Skew, Gaussian and Uniform.
%
%
The relative I/O cost will be defined in Section \ref{Experimental Setup}. Intuitively, if the relative I/O cost is smaller than 1, the index requires 
fewer nodes accesses than the R-Tree does. 
We observe from Table~\ref{Cost Function Based Action Space vs Child Node Based Action Space} that compared with the R-Tree, an RL model with the cost 
functions as actions only achieves an improvement of less than 2\% in terms of query processing time. 
We find from our experiments that in 90\% of the nodes, different cost functions end up with the same subtree choice which gives us very little room for improvement.

\begin{table}[h]
\normalsize
\vspace*{-3mm}
\centering
\caption{Performance of cost function based action space.}
\label{Cost Function Based Action Space vs Child Node Based Action Space}
\vspace*{-3mm}
\begin{tabular}{| c | c | c | c |}
\hline
& \multicolumn{3}{c|}{Relative I/O cost} \\ \hline
 & \textbf{Skew} & \textbf{Gaussian} & \textbf{Uniform} \\
\hline
\textbf{Use cost functions} & 0.98 & 0.98 & 1.00 \\
\hline
\textbf{Our final design} & 0.29 & 0.08 & 0.56 \\
\hline
\end{tabular}
\vspace*{-3mm}
\end{table}

As a result, we propose a new idea of 
training the RL agent to decide which child node to insert the new object directly.
Based on the idea, one design is to have all child nodes to comprise the action space. 
However, this incurs two challenges: 1) the number of child nodes contained by different nodes is usually different, and 2) the action space is large.
Considering all
child nodes as the actions leads to a large action space with many ``bad actions''.
The bad actions make the exploration during model training ineffective and 
inefficient. 
To address the challenges, we use the similar idea as we use for designing state space. 
%
Recall that in designing the state space, we propose to retrieve top-$k$ child nodes in terms of the 
increase of area to represent a state. To make the action space and the state
representation consistent, we define the action space $\mathcal{A}=\{1, \dots, 
k\}$, where action $a=i$ means the RL agent chooses the $i$-th retrieved child node
to be inserted with the new object.

\noindent\textbf{MDP: Transition.}
In the process of the \ChooseSubtree operation, given a state (a node in the R-Tree) and an
action (inserting the new object into a child node), the RL agent transits to
the child node. If the child node is a leaf node, the agent reaches a terminal 
state.

\noindent\textbf{MDP: Reward Signal.}
A reward associated with a transition corresponds to some feedback 
indicating the quality of the action taken at a given state. A larger reward 
indicates a better quality. Since our objective is to learn to build an R-Tree 
that processes query efficiently, the reward signal is expected to reflect the 
improvement of query performance.

In the process of \ChooseSubtree, it is challenging to directly evaluate if 
an action taken at a state is good, because the new object has not been
fully inserted into the tree yet. A straightforward idea is after the new object
has been inserted, we use the R-Tree to process a set of random range queries. The
inverse of the cost (e.g., the number of accessed nodes) for processing the queries 
is set as the reward shared by all of the state-action pairs encountered in the 
insertion of the new object. 
The agent seems to be encouraged to take the actions to build a tree that can process range queries by accessing as few nodes as possible. 
However, this is not the case due to the following reasons: (1) A previous action may
affect the tree structure and hence the 
query performance in the future. Therefore, a ``good'' action may receive a poor reward due to
some bad actions that were made previously. (2) More importantly, as we aim to 
learn to construct an R-Tree that outperforms the competitors, we are interested 
in knowing what kind of actions makes the resulting tree better than a 
competitor, and what kind of actions makes it worse. The aforementioned reward 
signal cannot distinguish the two types of actions, making it  ineffective 
for the agent to learn a good policy to outperform the competitors. (3) As more 
objects are inserted into the R-Tree, the average number of accessed nodes 
naturally increases. Therefore, the reward signal becomes weaker and weaker, 
which makes it difficult for the model to learn useful information in the late 
stage of the training.

Inspired by the observations, we design a novel reward signal for \ChooseSubtree. 
The high level idea is that we maintain a reference 
tree with a fixed \ChooseSubtree and \Split strategy. The reference
tree serves as a competitor and can be any existing R-Tree variant. The reward 
signal is computed based on the gap between costs for processing random queries 
with the reference tree and the RLR-Tree. Specifically, the design of the reward
signal is as follows:

(1) We maintain an R-Tree, namely RLR-Tree, that uses RL to decide which child
node to insert the new object, and adopts a pre-specified Split strategy.
	
(2) We maintain a reference tree which adopts a pre-specified \ChooseSubtree strategy and the same Split strategy as RLR-Tree.

(3) We synchronize the reference tree with the RLR-Tree, so that they have the 
	same tree structure.

(4) Given $p$ new objects $\{o_1, \dots, o_p\}$, we insert them into both 
	the reference tree and the RLR-Tree. 

(5) After the $p$ objects are inserted, we generate $p$ range queries of 
	predifined sizes whose centers are at the $p$ objects, respectively.

(6) The $p$ range queries are processed with both the reference tree and the 
	RLR-Tree. We compute the normalized node access rate, which is defined
	as $\frac{\text{\# acc. nodes}}{\text{Tree height}}$ and is the number of 
	accessed nodes for answering a range query over the tree height. Let $R$ and 
	$R'$ be the normalized node access rate of the RLR-Tree and the reference 
	tree, respectively. We compute $r = R' - R$ as the reward signal. The higher $r$ is, the fewer nodes 
	RLR-Tree needs to access to process the range queries than the 
	reference tree.

(7) All the transitions encountered in the insertion of the $p$ objects 
	share the same reward $r$.

With the 
idea, we are able to distinguish the good actions 
from the bad actions: A positive reward means that the RLR-Tree processes the 
recent $p$ insertions well as it requires fewer nodes accesses to process the 
queries compared with the reference tree. Moreover, as the reference tree is
periodically synchronized with the RLR-Tree, we can avoid the effect of previous 
actions. Therefore, maximizing the accumulated reward is equivalent to
encouraging the agent to take the actions that can make the RLR-Tree outperform 
the competitors.

\vspace{-2.5mm}
\subsubsection{Training the Agent for \ChooseSubtree}
\hfill

\noindent\textbf{Deep-$Q$-Network (DQN) Learning.}
Deep Q-learning is a commonly used model-free RL method. It uses a Q-function $Q^*(s, a)$ 
to represent the expected accumulated reward that the agent can obtain if it 
takes action $a$ in state $s$ and then follows the optimal policy until it 
reaches a terminal state. The optimal policy takes the action with the maximum $Q$-value
in any state. 
%
Deep-$Q$-Network \cite{mnih2015dqn} has 
been proposed to approximate the Q-function $Q^*(s, a)$ with a deep neural 
network $Q(s, a;\Theta)$ with parameters $\Theta$. In our model, we adopt the 
deep Q-learning with experience replay \cite{mnih2015dqn} for learning the $Q$-functions.

Given a batch of transitions $(s,a,r,s')$, parameters in $Q(s,a;\Theta)$ is updated 
with a gradient descent step by minimizing the mean square error (MSE) loss 
function, as shown in Equation~\ref{dqn_MSE}.

\begin{equation}
\vspace*{-2mm}
  L(\theta) = \sum_{s,a,r,s'}[(r + \gamma max_{a'}\hat{Q}(s',a';\Theta^-) - Q(s,a;\Theta))^2],
  \label{dqn_MSE}
\end{equation}

\noindent where $\gamma$ is the discount rate, and $\hat{Q}(;\Theta^-)$ is frozen target network.





\noindent\textbf{Training the Agent.}
We present the RL \ChooseSubtree training process in Algorithm~\ref{Insertion Model Training}.
We first initialize the 
main network $Q(s,a;\Theta)$ and the target network $\hat{Q}(s,a;\Theta^-)$ with 
the same random weights (line 3). In each epoch, it first resets the replay 
memory (line 5). Then it involves a sequence of insertions of the objects in the 
training dataset (lines 6--20). Specifically, for every $p$ objects $\{o_1, 
\dots, o_p\}$, we synchronize the structure of $T_r$ with $T_{rl}$ (line 7). For 
each $o_i$ of the $p$ objects, we first insert it into the reference tree (line 9). 
Then a top-down traversal on the RLR-Tree is conducted (lines 10--15). At each level,
we compute the state representation (line 12) and use $\epsilon$-greedy to choose
the action based on their $Q$-values (line 13), until we reach a terminal 
state (leaf node). The transitions are stored in $SA$ (line 14). At the leaf node, 
we insert the new object and use the same \Split strategy as the reference 
tree in a bottom-up scan to ensure no node overflows (line 15). Meanwhile, we 
generate a range query with a predefined size centered at $o_i$ and add the query 
to $RQ$ (line 16). When the $p$ objects have been inserted, we compute the reward 
with the queries in $RQ$ (line 17).
The reward computation process is illustrated in Figure \ref{RL ChooseSubtree Model Training}.
All transitions encountered in the insertions 
of the $p$ objects share the same reward $r$ and are pushed into the replay 
memory (line 18). Then we draw a batch of transitions randomly from the replay 
memory and use the batch to update the parameters in the main network $Q(;\Theta)$
as DQN does (line 19). The parameters in the target network $\hat{Q}$ are periodically 
synchronized with $Q$ (line 20). 

\begin{figure}[tbh]
\vspace*{-4mm}
\begin{center}
\noindent
  \includegraphics[width=\linewidth]{Figures/reward_computation.pdf}
  \end{center}
  \vspace*{-4mm}
    \caption{RL \ChooseSubtree Reward Computation}
    \label{RL ChooseSubtree Model Training}
\end{figure}

\noindent\textbf{Remark}. The new object to be
inserted may be fully contained in one of the child nodes. If we add the new object 
into such a child node, the MBRs of all child nodes are not affected. Therefore,
it is unnecessary to make the agent consider such cases. When such cases happen, we do not pass the state representation to the model, but 
choose the child node that contains the new object directly.


\vspace{-2mm}
\begin{algorithm}
\small{
  \SetAlgoLined
  \textbf{Input: }A training dataset\;
  \textbf{Output: }Learned action-value function $Q(s,a;\Theta)$\;
  Initialize $Q(s,a;\Theta)$, $\hat{Q}(s,a;\Theta^-)$\;
  \For{$epoch=1,2,\dots$}{
  	  Replay memory $\mathcal{M}\gets \emptyset$\;
  	  \For{every $p$ objects $\{o_1, \dots, o_p\}$ in dataset}{
  	  	$T_r \gets T_{rl}$, $SA\gets \emptyset$, $RQ\gets \emptyset$\;
  	  	\For{$o_i\in \{o_1, \dots, o_p\}$}{
  	  		Insert $o_i$ into $T_r$\;
  	  		$N\gets $ the root of $T_{rl}$\;
  	  		\While{$N$ is non-leaf}{
  	  		    $s\gets $ state representation of $N$ and $o_i$\;
  	  		    $a\gets $ an action selected by $\epsilon$-greedy based on $Q$-values\;
  	  		    $N\gets a$, $SA\gets SA\cup \{(s,a)\}$\;
  	  		}
	  	    Insert $o_i$ into $N$, split until no node overflows\;
	  	    $RQ\gets RQ\cup \{\text{a range query centered at } o_i\}$\;
  	  	}
  	  	$r\gets$ compute reward with queries in $RQ$\;
  	    Add $(s, a, r, s')$ for every $(s,a)\in SA$ into memory\;
  	    Draw samples from memory and update $\Theta$ in $Q(;\Theta)$\;
  	    Periodically synchronize $\hat{Q}(;\Theta^-)$ with $Q(;\Theta)$\;
  	  }
  } 
  \caption{DQN Learning for \ChooseSubtree}
  \label{Insertion Model Training}
  }
\end{algorithm}
\setlength{\textfloatsep}{0pt}

\subsubsection{Time Complexity}

In our analysis, the additional computation cost associated with the use of neural networks in an RLR-Tree is deemed constant. Assume the RLR-Tree has a size of $S$ and a height of $h$. Inserting an object into the RLR-Tree encounters $h-1$ states. At each state, it takes $O(k\cdot M)$ time to retrieve the top-$k$ child nodes and $O(M)$ to compute the features for each child node. Therefore, the overall time complexity is $O(h\cdotp k \cdot M)$. As a comparison, it takes $O(h\cdotp M)$ time for \ChooseSubtree in the R-Tree.

\vspace*{-2mm}
\subsection{Split}\label{sec:node_splitting}
The top-down traversal ends up at a leaf node. If the leaf node overflows, it will be split into 
two nodes and the \Split operation may be propagated upwards. Next, we present how to model \Split as an MDP and how to train the model.

\vspace*{-2mm}
\subsubsection{MDP Formulation}
We have also explored different ideas to design the state, action, transition and
reward signal of the MDP for \Split. 
Due to space limitation, we only present the final design here. 


\noindent\textbf{MDP: State Space.}
For \Split, it is natural that a state comes from an overflowing node. A straightforward idea is to make the representation of a state 
capture the goodness of all the possible splits, so that the agent can 
decide how to split the node. However, since an overflowing node contains 
$M+1$ entries, there are $(2^{M+1}-2)$ possible splits in total. It is 
impractical to reflect all of these splits in the state representation.


In order to avoid considering so many possible splits, we adopt a similar idea 
as R$^*$-Tree \cite{beckmann2009revisedrstartree} as follows: We first sort the 
entries with respect to their projection to each dimension. For each sorted 
sequence, we  consider the split at the $i$-th element ($m \leq i \leq M+1-m$), where 
the first $i$ entries are assigned to the first group and the remaining $M+1-i$ 
entries are assigned to the second group. This means for each sorted sequence, 
we only consider $M+2-2\cdot m$ splits. Next, we further discard the splits 
which create two nodes that overlap with each other. We pick the top-$k$ splits from the remaining 
splits in ascending order of total area and 
construct the representation of the state. Specifically, for each split, we 
consider four features: the areas and the perimeters of the two nodes created by 
the split. We concatenate the features of all $k$ splits and generate a 
$4k$-dimensional vector to represent the state. Note that the areas and 
perimeters are normalized by the maximum area and perimeter among all splits, so 
that each dimension of the state representation falls in $(0,1]$.

\noindent\textbf{MDP: Action Space.}
Similar to \ChooseSubtree, in order to make the action space consistent with
the candidate splits that are used to represent the state, we define the action 
space as $\mathcal{A}=\{1, \dots, 
k\}$, where $k$ is the number of splits used to 
represent the state. An action $a=i$ means that the $i$-th split is adopted.

\noindent\textbf{MDP: Transitions.}
In \Split, given a state (a node in the R-Tree) and an action (a 
possible split), the agent transits to the state that represents the parent
of the node. If the node does not overflow, it is the terminal state.


\noindent\textbf{MDP: Reward Signal.}
The reward signal for \Split is similar to that of \ChooseSubtree. We
maintain a reference tree that is periodically synchronized with the RLR-Tree.
We use the difference of the normalized node access rate as the two trees process training queries 
as the reward signal. Note that the RLR-Tree adopts the same 
\ChooseSubtree strategy as the reference tree and uses the RL agent to decide 
how to split an overflowing node.

\vspace{-1mm}
\setlength{\textfloatsep}{0pt}
\begin{algorithm}
\small{
  \SetAlgoLined
  \setcounter{AlgoLine}{0}
  \textbf{Input: }A training dataset\;
  \textbf{Output: }Learned action-value function $Q(s,a;\Theta)$\;
  
  Initialize $Q(s,a;\Theta)$, $\hat{Q}(s,a;\Theta^-)$\;
  \For{$epoch = 1,2,3, ...$}{
    \For{$j = 1,2,3, ... (parts-1)$}{
      training part $\gets \emptyset$\; 
      Construct $T_{base}$ with the first $\frac{j}{parts}$ of the objects (initial part)\;
      \For{$o$ in the remaining objects}{
      	\lIf{Inserting $o$ into $T_{base}$ causes a split}{Add $o$ to training part}
      	\lElse{Add $o$ to fill part and insert into $T_{base}$}
      }
      \For{every $p$ objects $\{o_1, \dots, o_p\} \subseteq$ training part}{
      	 $T_{rl}\gets T_{base}$, $T_{r}\gets T_{base}$, $SA\gets \emptyset$, $RQ\gets \emptyset$\;
      	\For{$o_i\in \{o_1, \dots, o_p\}$}{
      		Insert $o_i$ into $T_r$\;
      		Top-down traversal on $T_{rl}$ to a leaf node $N$\;
      		\lIf{$N$ overflows}{$RQ \gets RQ \cup \{\text{training query}\}$}
      		\While{$N$ overflows}{
      		    $s\gets$ state representation of $N$\;
      		    $a\gets$ an action selected by $\epsilon$-greedy based $Q$-values\;
      		    $N\gets N$'s parent, $SA\gets SA\cup \{(s,a)\}$
      		}
      	}
      	$r\gets$ compute reward with queries in $RQ$\;
      	Add $(s,a,r,s')$ for all $(s,a)\in SA$ to  memory\;
      	Draw samples from memory and update $Q(;\Theta)$\;
      	Periodically synchronize $\hat{Q}(;\Theta^-)$ with $Q(;\Theta)$\;
      }
      
    }
   }
  \caption{DQN Learning for \Split}
  \label{Splitting Model Training}
}
\end{algorithm}
\setlength{\textfloatsep}{0pt}

\vspace*{-4mm}
\subsubsection{Training the Agent for Split}
Compared with \ChooseSubtree, training the agent for \Split is a more 
challenging task. 
To insert an object into the R-Tree, \ChooseSubtree is iteratively invoked at each level in the top-down traversal. On the other hand, \Split operation is invoked only when a node overflows. Therefore, only a few transitions for
\Split are available for training, making it difficult for 
the agent to learn a good policy in reasonable time.

To tackle this challenge, we design a new method for the agent to 
interact with the environment, so that \Split operation is frequently invoked.
Theoretically, if all the nodes are
full, the insertion of a new object definitely causes a tree node to split. 
Inspired by this, we propose to first build a tree in which most of the nodes
are full, so that node splits can be frequently encountered. We generate such 
R-Trees with different sizes, and use the transitions caused
by inserting the remaining objects in the dataset for training.

The procedure is presented in Algorithm~\ref{Splitting Model Training}. 
In each epoch, we repeat $parts-1$ iterations to train
the agent (lines 5--24). In particular, in the $i$-th iteration, the first
$\frac{i}{parts}$ of the training dataset forms the initial part which is used to construct an R-Tree 
$T_{base}$ (line 7). The remaining data are then divided into 2 parts, i.e., the fill part containing objects that will not cause node overflow in $T_{base}$ and the training part containing objects not in the fill part. Objects in the fill part are inserted into 
$T_{base}$ while objects in the training part will be used to trigger
splits for training later (line 9--10). 
In this way, most of the nodes in $T_{base}$ are likely to be full and the objects in the training part are likely to cause splits. This pre-training preparation process is illustrated in Figure \ref{RL Split Model Training}. After $T_{base}$ 
is constructed, we start training with objects in the training part (lines 11--24). 
For every $p$ objects, we first synchronize $T_r$ and $T_{rl}$ with $T_{base}$ 
(line 12). This makes $T_r$ and $T_{rl}$ have the same structure and are almost 
full. Then for each of the $p$ objects, we insert it into the reference tree 
$T_r$ directly with pre-specified \ChooseSubtree and \Split strategies (line 14).
For the RLR-Tree, we use the same \ChooseSubtree strategy as the reference tree to
reach a leaf node $N$ (line 15). If $N$ overflows, we 
generate a range query with a predefined size centered at $o_i$ and add the query 
to $RQ$ (line 16). Then we iteratively split $N$ and move to its parent until 
$N$ does not overflow (lines 17--20). For each node $N$, we compute the state
representation and use $\epsilon$-greedy to select an action based on their 
$Q$-values (lines 18--19). The transitions are stored in $SA$ (line 20). When
the $p$ objects have been processed, we compute the reward with the queries in
$RQ$ (line 21). The transitions encountered in processing the $p$ objects share
the same reward (line 22). Then we draw a batch of random transitions from the
replay memory and use it to update the parameters in the main network (line 23).
The parameters in the target network $\hat{Q}(;\Theta^-)$ are periodically 
synchronized with the main network $Q(;\Theta)$ (line~24). 

\begin{figure}[tbh]
\vspace*{-4mm}
\begin{center}
\noindent
  \includegraphics[width=\linewidth]{Figures/construct_base_tree.pdf}
  \end{center}
  \vspace*{-4mm}
    \caption{RL \Split Pre-Training Preparation}
    \label{RL Split Model Training}
\vspace*{-4mm}
\end{figure}

\noindent\textbf{Remark.} When we split a node to two, the ideal case is that the two nodes do not overlap with each other, so that the number of node accesses
for processing a query can be reduced. It is more challenging when there are many
candidate splits that generate two nodes with zero overlap, as we need to carefully
consider how to break the tie. As a result, we consider such a special case in the 
exploration of the agent. Specifically, if there exists at most one candidate split
that generates two non-overlapping nodes, we simply select the split with the 
minimum overlap. We only use RL to decide how to split the node when more
than one split generates non-overlapping nodes. In this way,
the agent is trained more effectively.

\smallskip
The training process for \Split has two advantages. Firstly, using
different fractions of objects to build an ``almost-full'' base tree helps to
learn a more general model. Secondly, by building the ``almost-full'' base tree
and periodically resetting $T_r$ and $T_{rl}$ to the base tree ensures that the
\Split operation is consistently invoked at a high frequency. This makes the 
training process more efficient.

\vspace{-2mm}
\subsubsection{Time Complexity}
Similar to \ChooseSubtree, the additional computation cost associated with the use of neural networks is deemed constant. If a node overflows, at most $O(h)$ Split operations are invoked.
In each Split operation, we first sort the entries along each dimension, which takes $O(M\log M)$ time. Then it takes $O(k\cdotp (M-2m)\cdotp M)$ time to retrieve the top $k$ splits with minimum total area. Finally, it takes $O(k\cdotp M)$ time to compute the four features for the $k$ splits. Therefore, the overall time complexity for RLR-Tree \Split is $O(h\cdotp k\cdotp (M-2m)\cdotp M)$. As a comparison, \Split operation takes $O(h\cdotp M^2)$ time in the classic R-Tree.

\vspace{-3mm}
\subsection{The Combined Model} 
\label{Combination of Insertion and Splitting Models}

We expect that the two agents are closely tied together and are able to help each other achieve better performance. 
%
%
A possible idea for combing the two models is to make the agents share information such as the states observed and the corresponding actions selected with each other. Since both RL models have the same goal, this kind of information sharing has the potential to enhance the agents' policy learning. However, it is difficult for the two models to exchange information, as they are dealing with different procedures. This approach does not lead to any query performance improvement in the final RLR-Tree. 

Recall that we specially design a learning process of Split, as node overflow occurs infrequently in the construction of an R-Tree. 
Motivated by this, 
we propose an enhanced training process to train the two agents alternately. 
Specifically, in odd epochs, we train the RL agent for \ChooseSubtree
and the agent for \Split is fixed to be the \Split strategy for the RLR-Tree.
In even epochs, we train the agent for \Split and the agent 
for \ChooseSubtree is fixed to be the \ChooseSubtree strategy for RLR-Tree.

\subsection{RLR-Tree Construction \& Dynamic Updates}
\label{construction}

With the learned models for \ChooseSubtree and \Split, we incorporate the models into the insertion algorithm of R-Tree to build the RLR-Tree as follows.
%
%
For each object to be inserted, in the top-down traversal, we iteratively compute the state representation of a node and use the model trained for \ChooseSubtree to select the subtree corresponding to the action with the maximum $Q$-value until the object is inserted into a leaf node. When a node overflows, we compute its state representation and use the model trained for \Split to choose the split corresponding to the action with the maximum $Q$-value.
For dynamic updates, new data records can be inserted into an existing tree using the trained models. Our experimental results (Section \ref{dynamic updates}) show that the trained models do not experience obvious performance deterioration even when there is a change in data distribution.